\documentclass[%
 reprint,
superscriptaddress,
nofootinbib,
 amsmath,amssymb,
 aps,
showkey,
notitlepage,10pt
]{revtex4-1}
\usepackage{graphicx}
\usepackage{etoolbox}
\usepackage{scalerel}
\usepackage{cleveref}
\usepackage{makecell}
\usepackage{braket}
\usepackage[table]{xcolor}
\usepackage[final]{changes}
\definecolor{gray2}{rgb}{0.8,0.8,0.8}

\newcommand{\Tbath}{\ensuremath{T_{\rm{b}}}}

\newcommand{\aalto}{QTF Centre of Excellence, Department of Applied Physics,\\ Aalto University, P.O. Box 13500, FI-00076, Aalto, Finland}
\newcommand{\unsw}{School of Electrical Engineering and Telecommunications,\\
University of New South Wales, Sydney, New South Wales 2052, Australia}

\newcommand{\npl}{National Physical Laboratory, Hampton Road, Teddington TW11 0LW, United Kingdom}
\newcommand{\strathclyde}{Department of Physics, SUPA, University of Strathclyde, Glasgow G4 0NG, United Kingdom}
\newcommand{\vtt}{VTT Technical Research Centre of Finland Ltd, P.O. Box 1000, FI-02044, VTT, Finland}

\begin{document}
\title{\Large{Superconducting charge sensor coupled to an electron layer in silicon}}

\author{M\'at\'e Jenei}
\email[]{mate.jenei@aalto.fi}
\affiliation{\aalto}
\author{Ruichen Zhao}
\thanks{ Current address: NIST 325 Broadway, Boulder, CO 80305, USA}
\affiliation{\unsw}
\author{Kuan Y. Tan}
\affiliation{\aalto}
\affiliation{\unsw}
\author{Tuomo Tanttu}
\affiliation{\unsw}
\author{Kok W. Chan}
\affiliation{\unsw}
\author{Yuxin Sun}
\affiliation{\unsw}
\author{Vasilii Sevriuk}
\affiliation{\aalto}
\author{Fay Hudson}
\affiliation{\unsw}
\author{Alessandro Rossi}
\affiliation{\strathclyde}
\affiliation{\npl}
\author{Andrew Dzurak}
\affiliation{\unsw}
\author{Mikko M\"ott\"onen}
\affiliation{\aalto}
\affiliation{\vtt}

\begin{abstract}
Schemes aimed at transferring individual electrons in semiconductor devices and detecting possible transfer errors have increasing importance for metrological applications. 
We study the coupling of a superconducting Josephson-junction-based charge detector to an electron island defined by field-effect in silicon. The flexibility of our device allows one to tune the coupling using the detector as an additional gate electrode. We study the reliability of the electron sensor in different device configurations and the suitability of various operation modes for error detection in electron pumping experiments. As a result, we obtain a charge detection bandwidth of 5.87 kHz with unity signal to noise ratio at 300 mK bath temperature. 
\end{abstract}

{\let\newpage\relax\maketitle}

\section{Introduction}

In May 2019, the unit ampere was redefined by linking it to an exact numerical value of the elementary charge \cite{BIMP9th}. This has reinforced the need for practical experimental methods that allow one to generate a macroscopic electric current by controlling electron transfer at the single-particle level. Single-electron (SE) current sources that have proved successful in generating sufficiently high current values, hundreds of pA, with sub-ppm uncertainty are based on semiconductor quantum dots (QDs) \cite{Angus2007, Lim2009, Giblin2019}. These devices are commonly referred to as SE pumps because they operate by transferring a fixed number of charges, $n$, at a rate given by the frequency, $f_\textup{p}$, of a periodic drive signal applied to at least one of the gates of the device \cite{Pekola2013,Tanttu2016}. The resulting direct electric current is \mbox{$I_\textup{p}=nef_\textup{p}$}, where $e$ is the elementary charge. 

The accuracy of a pump is ultimately given by the mean number of electrons transferred per cycle of the periodic drive. Errors in the pumping protocol may lead to excess or missed transitions with respect to the ideal value of $n$. A number of experimental works have demonstrated the detection of such errors in real time with charge detectors capacitively coupled either to a counting island (CI) that collects pumped electrons \cite{Keller1996,Fricke2014,Yamahata2014a} or to the QD directly \cite{Giblin2016}. 

The first error counting experiment employed a metallic pump, which shuttled electrons with a frequency of 5.05 MHz \cite{Keller1996}. In contrast, GaAs-based SE pumps transfer charges in the range of hundreds of MHz and the counting circuits operate in the range of a hertz in a strong magnetic field which is compatible with non-superconducting detectors \cite{Fricke2013,Fricke2014,Giblin2016}. Silicon pumps, however, require only electrostatic confinement to achieve single-electron pumping at GHz frequencies \cite{Rossi2014,Yamahata2016,Zhao2017}, but silicon-based detectors provide limited sensitivity and counting bandwidth \cite{Yamahata2014a,Tanttu2015}. In contrast, superconducting single-electron transistor (SSET) charge sensors have higher sensitivity and lower back action \cite{Xue2009,Cassidy2007,Lu2003RealtimeDO,Yuan2011}. Metallic SETs, for charge detection, provide more flexibility also compared with quantum point contacts because of the possibility to tune both the detector coupling strength to the CI \cite{Sun2009} and the charge sensitivity \cite{Kuzmin1989}. 

In this work, we study a double-Josephson-junction-based SSET charge detector, which is capacitively coupled to a CI in a two-dimensional electron gas (2DEG). We focus on investigating the two operation modes of the detector, namely the standard coupling, in which the metallic detector directly couples to the CI, and the enhanced coupling, in which the SSET induces an intermediate charge island to increase the charge sensitivity. We report the charge and time stability of the detector, the coupling between the sensor and the CI, and measure the detector noise in both of the operation modes. As a conclusion, we find that the SSET charge detector in the standard operation mode has lower charge noise and higher coupling compared with a Si-SET in a similar circuit \cite{Tanttu2015}, which results at least a factor of 90 improvement in the detection bandwidth.

\begin{figure}
\centering
\includegraphics[width=\linewidth]{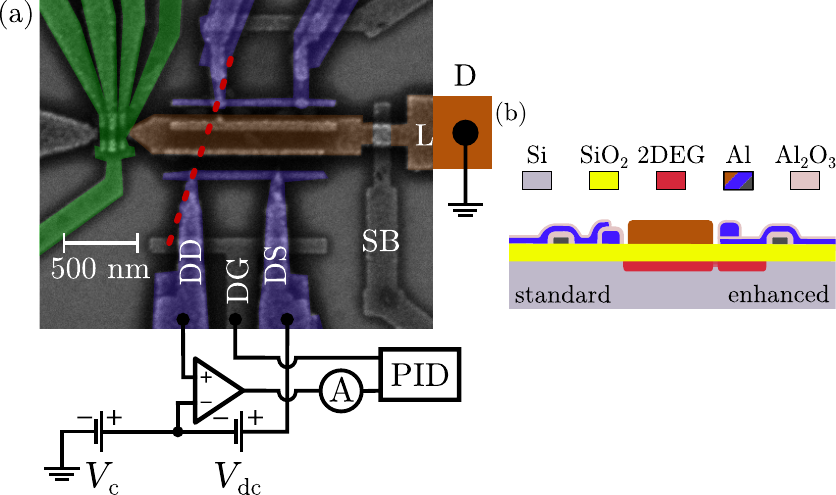}
\caption{(a) False-color micrograph of a device and the setup displaying the charge detectors (blue), the electron pump (green), and the counting island (orange). (b) Schematic cross-section of the device along the red dashed line in panel (a) indicating the standard (left detector) and enhanced (right detector) coupling modes.}
\label{fig:figure1}
\end{figure}

\section{Device architecture}

A device identical to that used in the experiments is depicted in Figs.~\ref{fig:figure1}(a) and \ref{fig:figure1}(b). Two SSETs are patterned on each side of a CI that is induced under a lead gate (L) by applying a positive gate voltage $V_{\rm{L}}$. Each SSET is contacted by a pair of superconducting electrodes (DS and DD). To tune the detector sensitivity, we use an auxiliary detector gate (DG). Operating the device with two detectors simultaneously provides more statistics on the state of the CI, but we use only the bottom detector which has lower normal-state resistance, $R_{\mathrm{J}}$, than the top detector. The quantum dot pump is located on the left side of the CI but it has not been used in this experiment. On the right, there is a cryogenic switch (SB) which controls the galvanic coupling between the CI and an ohmic drain (D).

The samples are metal-oxide-semiconductor (MOS) nanostructures, i.e., a stack of Al/Al$_2$O$_3$ electrodes fabricated on intrinsic silicon with a thermally-grown 8-nm-thick SiO$_2$ gate oxide. The device metallization consists of two parts. First, the gates of the pump and the CI are defined with 3 steps of electron beam lithography, metal evaporation and lift-off. Next, the SSETs are deposited with a double-angle evaporation technique, where the two depositions are separated by an \textit{in-situ} oxidation layer.

The capacitive coupling between the SSET and the CI can be increased by inducing an intermediate island beneath the SSET detector island \cite{Sun2009}, thus changing the detection from standard coupling to enhanced coupling. By operating gate DG below or near the threshold voltage, one can control the extension of the additional 2DEG. Experimentally, we vary a common voltage $V_\mathrm{c}$ between the SSET and the ground. We carried out numerical simulations that suggest a factor of ten increment in the sensitivity in the enhanced-coupling regime compared with the standard coupling (see Appendix \ref{appendix:simulation}). Our experiments focus on the mutual properties of the detector and the CI in the stationary case. Charge pumping is not studied in this work. 

\begin{figure*}
\centering
\includegraphics[width=\linewidth]{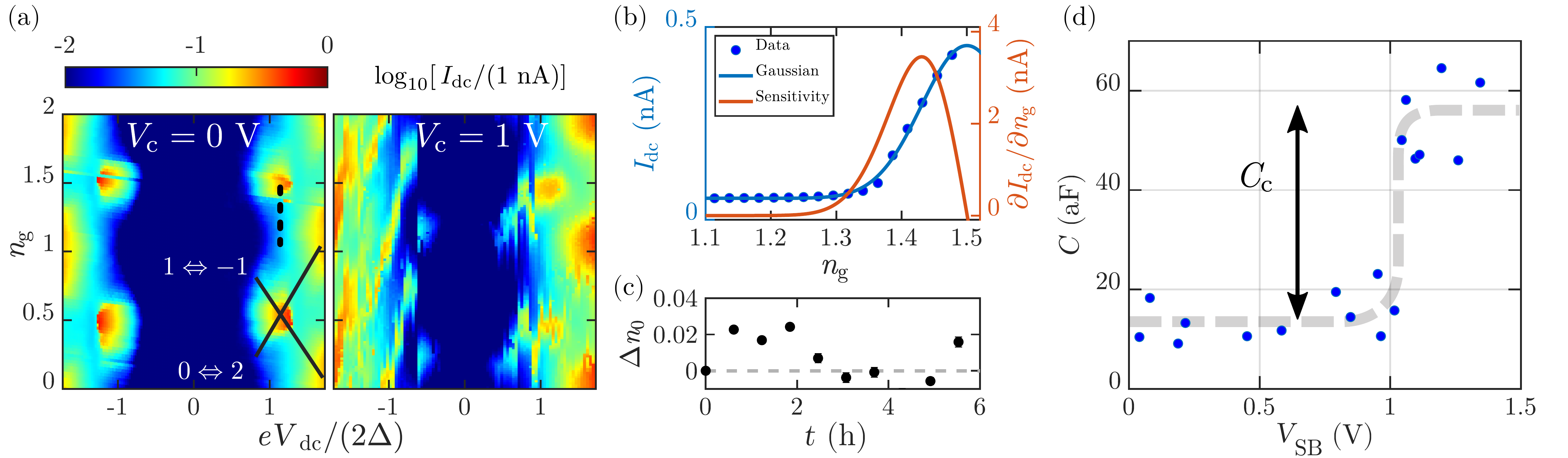}
\caption{(a) Electric current through the detector as a function of the bias voltage and gate charge in standard coupling (left) and in enhanced coupling configuration (right). (b) Current-to-charge conversion along the dashed line in panel (a). The sensitivity of the detector (red line) is extracted from the Gaussian profile of the current-to-charge conversion (blue line). (c) Offset charge stability at the bias voltage corresponding to the DJQP. (d) Capacitive coupling between the detector and the counting island as the function of SB voltage (blue dots) in the standard-coupling scheme, where $V_{\rm{L}} = 1.5 ~$V. The gray dashed line is a guide to the eye.}
\label{fig:figure2}
\end{figure*}

\section{Analysis and results}
\subsection{Coulomb stability}

When the detector is operated in a voltage-biased mode, a constant electric current flows through it until a change in the local electrostatic field, arising from an electron tunneling, changes the conductance through the detector. The voltage on DG varies the gate-induced quasiparticle number $n_{\rm{g}}$ on the detector island, which tends to change the total number of electrons, $n$, on the island and creates a periodic feature in the detector current $I_{\rm{dc}}$ \cite{Kuzmin1989}. Any electrostatic disturbance of the detector appears as current noise, $\delta I_{\rm{dc}}$, which is converted to charge fluctuation using
\begin{equation}
\delta n_{\mathrm{g}} =  \left(\frac{\partial I_{\mathrm{dc}}}{\partial n_{\mathrm{g}}}\right)^{-1} \delta I_{\mathrm{dc}},
\label{eq:gate_charge_to_current_conversion}
\end{equation}
where $\partial I_{\mathrm{dc}}/\partial n_{\mathrm{g}}$ is the SSET sensitivity. Since we aim to observe single-electron events, all quantities are expressed in the units of $1e$, i.e., the electric current shift caused by an addition of a quasiparticle on the detector island. 

We utilize the DG to record the Coulomb stability diagram of the device. The resulting direct current (dc) of the detector as a function of $n_{\rm{g}}$ and the dc bias voltage $V_{\rm{dc}}$ is presented in \mbox{Fig.~\ref{fig:figure2}(a)}. We use a room temperature transimpedance amplifier (Femto DDPCA-300) to measure the electric current through the detector. At a certain $n_{\mathrm{g}}$, the electrochemical potential of the left (right) electrode equals that as the detector island, a supercurrent appears that matches with the $n = 0 \Leftrightarrow 2$ ($n = -1 \Leftrightarrow 1$) charge transition. Along such Cooper pair resonance lines, the quasiparticle number on the island remains constant. At $eV_{\mathrm{dc}}= 2E_{\mathrm{c}}$ and $n_{\mathrm{g}}=1/2$, where $E_{\mathrm{c}}$ is the charging energy of the SSET, the crossing of the two resonant Cooper-pair features determines the location of the 1$e$-periodic double Josephson quasiparticle process (DJQP) \cite{vanderBrink1991,Clerk2002,Lu2002}, which assigns our operation point, where the detection may reach a quantum limit of efficiency due to suppressed backaction \cite{Clerk2010a} which has been demonstrated in rf experiments \cite{Thalakulam2004,Xue2009}. From the location of the DJQP and the high-bias Coulomb diamonds (not shown), we extract \mbox{$E_{\rm{c}} = 245~\mu$eV}, $R_{\rm{J}} = 110 $ k$\Omega$, and the superconducting gap $\Delta = 200~\mu$eV. 
When the detector is tuned to enhanced coupling, \mbox{$V_{\rm{c}} = 1 $ V}, the DJQPs maintain their location and hence the shift in the charging energy is negligible. From the change of the gate voltage periodicity, on the other hand, the detector-island-to-DG coupling decreased from $17.5~$aF to $16~$aF. Because of the hindered SSET stability at $V_{\rm{dc}}< 0 $ V, we use the positive bias point. The current-to-charge conversion is extracted along the DJQP in Fig. \ref{fig:figure2}(b). Due to thermal broadening, the expected Lorenzian shape \cite{Clerk2002} is deformed into a Gaussian in the vicinity of the operation point.

\subsection{Detector charge drift}
To quantify the charge drift in the detector in the standard coupling scheme, we measure the Coulomb oscillations at the bias voltage point corresponding to the DJQP. The range in the gate voltage covers 6 resonances that have been measured in positive and negative sweep directions to observe and compensate for any hysteresis. The extracted drift of the offset gate charge is shown in Fig. \ref{fig:figure2}(c). The DJQP offset charge stability is extracted from the shift of the half-maximum direct current of the Coulomb oscillation and converted to gate charge using Eq.~\eqref{eq:gate_charge_to_current_conversion}.
The offset drift is within $|\Delta n_{\mathrm{0}}(t)| <  3\times10^{-2} $ over six hours, which is in good agreement with previous SSET experiments \cite{Zimmerman2008,Zimmerman2014}. However, complementary metal oxide semiconductor devices may exhibit very low 1/$f$ noise with $|\Delta n_0\mathrm{(t)}| \leq 10^{-2}$ drift observed over eight days \cite{Koppinen2013}. 

\subsection{Coupling between the counting island and the detector  }

The capacitive coupling between the CI and the detector  $C_{\mathrm{c}}$ is extracted using a lock-in technique detailed in Appendix \ref{appendix:coupling}. This measurement has been carried out in the standard-coupling mode in a device, which has an identical geometry to that used to produce the rest of the results. The sinusoidal output of the lock-in amplifier is connected to the drain of the device, while the bias voltage on SB controls the conductance between the CI and the drain electrode. The ac and dc components of the electric current are simultaneously measured at the SSET drain with a lock-in amplifier and a digital multimeter, respectively.
In this configuration, the capacitance between the drain electrode and the SSET is extracted

\begin{equation}
C =  \frac{en^{\mathrm{rms}}(V_{\mathrm{SB}})}{V_{\mathrm{ac}}^{\mathrm{rms}}},
\label{eq:coupling}
\end{equation}
where the $n^{\mathrm{rms}}$ is the measured root-mean-square (rms) amplitude of the induced gate charge and $V_{\mathrm{ac}}^{\mathrm{rms}}$ is the rms voltage amplitude excitation from the lock-in amplifier.
If the $V_{\mathrm{SB}}$ is low, the drain electrode is essentially decoupled from the CI. Tuning the SB voltage above the threshold couples the rms excitation in the 2DEG to the detector as Fig.~\ref{fig:figure2}(d) indicates. At $V_{\mathrm{SB}}= 0$ V, finite coupling is present between the drain and detector due to remaining stray coupling. 

Using \mbox{Eq. \eqref{eq:coupling}}, we find a 45-aF coupling between the CI and the detector island. The corresponding variation in the gate charge from adding a single electron at the CI, i.e., the charge sensitivity, is  ${\delta q_{\mathrm{e}}} = \kappa e,$ where $\kappa = \frac{C_\mathrm{c}}{C_{\mathrm{\Sigma_{CI}}}}$ and $C_{\mathrm{\Sigma_{CI}}} \simeq 1.9$ fF is the estimated total capacitance of the CI using a parallel-plate-capacitor approximation. We obtain ${\delta q_{\mathrm{e}}} = 2.36\times 10^{-2} ~ e$ that is comparable with our numerical result ${\delta q_{\mathrm{e}}} = 2 \times 10^{-2}$ $e$. However, in the enhanced-coupling mode the expected charge sensitivity is ${\delta q_{\mathrm{e}}} = 2.15 \times 10^{-1} ~e$. Unfortunately, the measurement in the enhanced coupling is hampered by random switching events described in \mbox{Sec. \ref{subsect:jump_noise}}. 
\begin{figure*}
\centering
 \includegraphics[width = \linewidth]{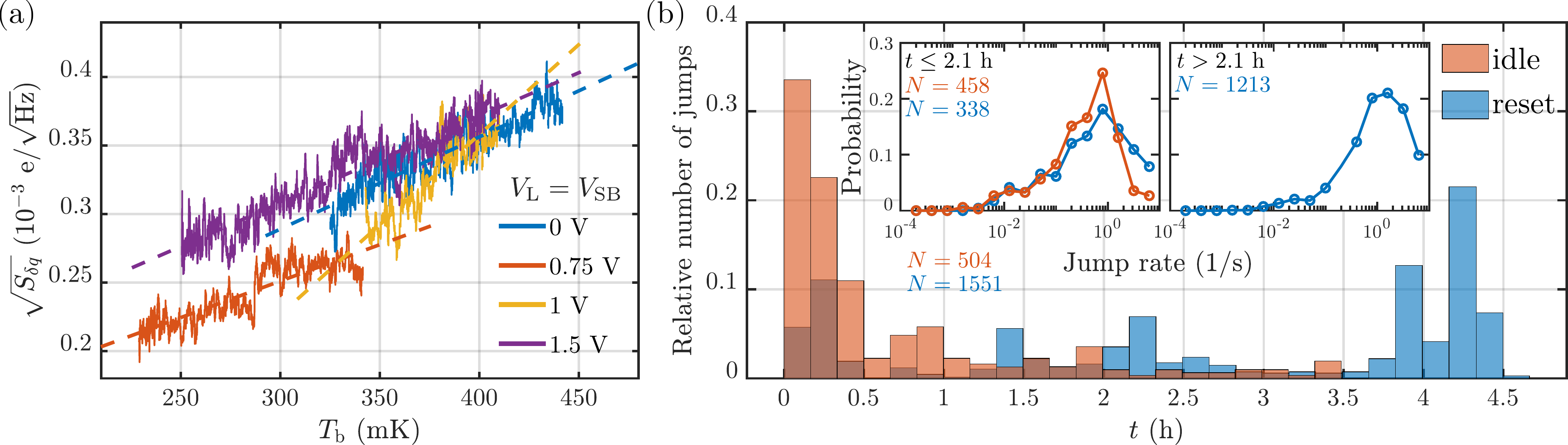}

        \caption{(a) White-noise level of the SSET in the standard-coupling mode during temperature drift at different voltages applied on the L and SB gates. The charge noise is calculated form a moving 80-s-long time window of data (solid lines) used to fit the linear temperature dependence of the noise (dashed lines). (b) Jump noise in the enhanced-coupling mode. The relative number of jumps in the idle (red) and reset (blue) states presented in 10-min-long segments. The inset shows the probability distribution of the jump rates for the approximately first half of the total measurement time (left) and for the second half (right). Here, $N$ denotes the number of jumps in the corresponding sections.}
        \label{fig:figure3} 
\end{figure*}

\subsection{Charge noise on the detector}
\label{subsect:change_noise}
 The output of the room temperature amplifier is connected to a proportional-integral-derivative (PID) controller that tunes the DG voltage to maintain the SSET sensitivity constant during the experiments and to compensate the 1/$f$ noise that is caused presumably by background charge motion.
 Although the highest sensitivity is desirable, the charge noise may cause the PID controller to jump to an adjacent Coulomb peak, i.e., to load or remove a electron from the detector island. Hence, we have to make a compromise between sensitivity and stability, and hence choose a setpoint current on the positive slope of the current-to-charge curve that is below the value that corresponds to the most sensitive point.

First, the SSET noise is characterized at standard coupling, as shown in Fig. \ref{fig:figure3}(a). To compensate for the drifts in the bath temperature $\Tbath$, we record for each experiment the evolution of the standard deviation of the detector gate charge, calculated in a 80-s-long time window, to quantify the noise. During the measurement, $\Tbath$ changes approximately linearly in time. Subsequently, the temperature-dependent charge noise, $\sqrt{S_{\delta q}}$, is determined by dividing the measured standard deviation by the square root of the measurement bandwidth. The fitted linear dependence between $\Tbath$ and the charge noise serves to indicate the dominance of the white noise in our detector, as expected in the high-temperature regime.

Similarly to the above-mentioned method, we vary the voltage on the L and SB gates and measure the detector noise. The gate voltages are chosen to be at the turn-on voltage of the gates  $0.75 $ V, in the beginning of the saturation $1 $ V, and above the saturation voltage $1.5 $ V. In all three cases, the charge noise depends linearly on $\Tbath$, as shown in \mbox{Fig.~\ref{fig:figure3}(a)}, but the slopes and the offsets are different. Consequently, the white-noise properties are sensitive to the local electrostatic field configuration. Also step-like variations appear in the charge noise measurement for the voltage configurations \mbox{$0.75 $ V} and $1.5 $ V at \mbox{$\Tbath = 280 $ mK} and \mbox{$\Tbath = 340$ mK}, respectively.
\subsection{Jump noise in enhanced coupling regime}
\label{subsect:jump_noise}
In the enhanced-coupling mode, $V_\mathrm{c}= 1$ V, large-amplitude jumps in the detector signal appear. The island underneath the metallic detector presumably activates and amplifies the impact of the switching events of two-level fluctuators in the vicinity of the extended CI \cite{Furlan2003,Pourkabirian2014}, which hinders the above-discussed estimation of the charge noise and interferes with the charge detection. The time traces exemplifying the jumps are presented in Appendix \ref{appendix:jumps}. We characterize the transient jumps by monitoring the detector after creating the CI, \mbox{$ V_{\rm{L}} = 1.5 $ V,} and we count the events, for which the detector falls out of the 3.5$\times\sigma$ white-noise level. The relative number of jumps are counted in 10-min-long segments for 4.5 hours, which are normalized by the total number of jumps, presented in \mbox{Fig. \ref{fig:figure3}(b)}. First, the system is measured in an idle state, where $V_{\rm{SB}} = 0 $ V and hence the CI is not tunnel coupled to the drain. Then we connect the CI to the drain electrode by applying a dc voltage $V_{\rm{SB}} = 1.5 $ V. The transparent SB barrier evacuates any excess charge from the CI, and hence resets the system. In the idle state, the number of jumps decreases with time, essentially vanishing after 3.5 hours from the introduction of the CI. On the other hand, the reset state exhibits an increasing number of jumps after 3.5 hours.

In addition to the relative number of switching, the jump rate, i.e., the inverse waiting time between consecutive jumps is an important quantity. The probability distributions of the jump rates are shown in the insets of \mbox{Fig. \ref{fig:figure3}(b)}. The first 2.1 hours are mostly dominated by jumps at approximately every second for both reset and idle states. After 2.1 hours in the reset state, the jump rates resemble the first 2.1 hours statistics. Because in the idle state the number of events is too small, the jump rate distribution in this configuration is omitted. In the enhanced-coupling mode, such statistics may help to filter false-positive charge detection events by defining time scales for error counting.

\subsection{Application in metrology}
To study the charge transport dynamics and the theoretically-predicted pumping accuracy using our detector, we estimate what is the lowest pumping performance that can be resolved. The SSET couples to the CI, which responds to failed pumping events occurring with a rate $\Gamma_{\mathrm{err}}$ \cite{Scherer2017}. The required averaging time to detect a single-electron event on the CI with signal-to-noise ratio of unity is 
\begin{equation}
t_{\mathrm{det}}= \frac{1}{2} \frac{S_{\delta q}}{{\delta q_{\mathrm{e}}}^2}.
\label{eq:det_time}
\end{equation}
In standard coupling, when \mbox{$V_{\rm{L}} = 1.5 $ V}, the extracted noise level is \mbox{$S_{\delta q} ^{\frac{1}{2}} =  3.08 \times 10^{-4} \ e/\sqrt{\mathrm{Hz}} $} at $\Tbath = 300 $ mK and ${\delta q_{\mathrm{e}}} = 2.36\times 10^{-2} ~ e$. Using Eq. (\ref{eq:det_time}), we obtain the detection time  $t_{\mathrm{det}} = $ 85 $\mu$s.

Thus the upper bound for the error rate in the single-electron pump which the detector resolves is \mbox{$\Gamma_{\mathrm{err}} =5.87 \times 10^3~$1/s}. Therefore, the worst pumping performance which can be measured with the detector at \mbox{1 GHz} is \mbox{5.87 ppm}. For silicon quantum dot pumps, a lower bound of \mbox{4 ppb} for the pumping uncertainty at \mbox{1 GHz} has been predicted in Ref.~\cite{Zhao2017}, taking only thermal effects into account. However, non-adiabatic excitations arising from the rf driving has an important role in waveform-optimized pumps, which has been considered to induce pumping errors in the range of \mbox{10 ppb} \cite{Kataoka2011,Rossi2014,Kashcheyevs2010,Waldie2015,Kaestner2015}. 
 
\section{Conclusion}
In conclusion, our results demonstrate the potential of a superconductor-based charge detector for future error detection experiments in silicon single-electron pumps. Namely, we observed low white-noise levels, high tunability, and high charge sensitivity.  
We have studied the charge stability of our charge detector, highlighting possible issues in the enhanced-coupling regime, namely the high-amplitude jumps in the dc current signal. The white-noise level and charge sensitivity in the standard-coupling mode demonstrates a significant advancement over recent charge counting experiments in silicon \cite{Tanttu2015}. According to the measured white-noise level and detector-to-CI coupling, future error counting experiments in silicon devices are feasible within a \mbox{5.87-ppm} relative uncertainty of the pumped current to benchmark the electron transfer at 1 GHz. Thus our work could enable high accuracy diagnostics for single-electron transport dynamics and metrological applications.

\begin{acknowledgments}
This work was carried out as a part of the Academy of Finland Centre of Excellence program (312300) and Academy of Finland projects 308161, 314302, and 316551. We acknowledge the provision of facilities and technical support by Aalto University at OtaNano - Micronova and the NSW Node of the Australian National Fabrication Facility, where the devices were fabricated. We acknowledge funding from the Joint Research Project ‘e-SI-Amp’ (15SIB08) and the Australian Research Council project DP160104923. This project has received funding from the European Metrology Programme for Innovation and Research (EMPIR) co-financed by the Participating States and from the European Union Horizon 2020 research and innovation programme.   
\end{acknowledgments}

\nocite{*}
\appendix

\section{Numerical Simulation}
\label{appendix:simulation}
\begin{figure}[ht]
\centering
\includegraphics[width=\linewidth]{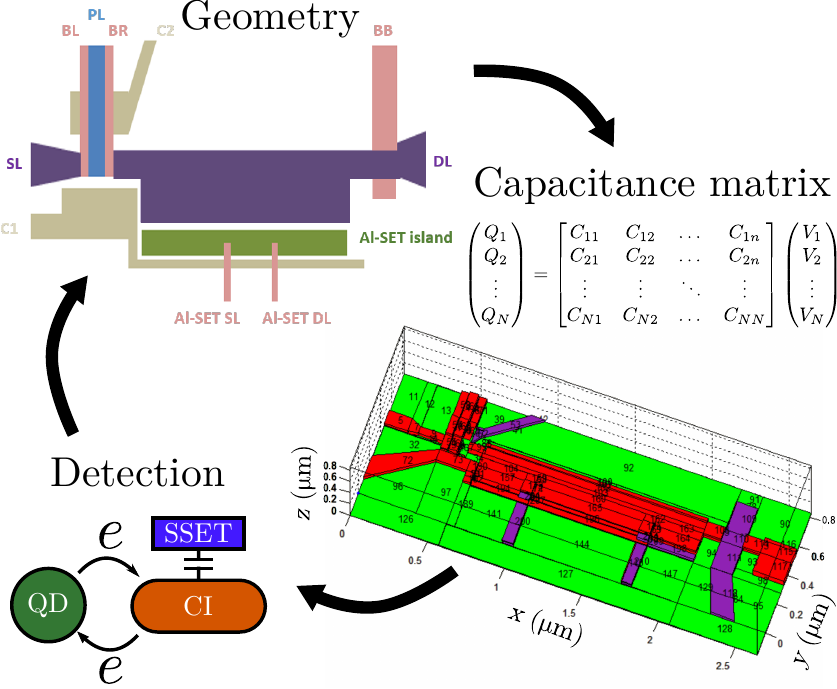}
\caption{Iterative steps of the numerical simulation. First, the device layout is designed including all the nanosize features. In the next step, the total capacitance matrix is calculated based on a finite-element simulation, which we use in the consecutive step to simulate the charge detection experiment. The geometry is then modified and the process is reiterated.} 
\label{fig:appfigure3}
\end{figure}

To characterize the device design, we employ an iterative numerical simulation technique presented schematically \mbox{in Fig. \ref{fig:appfigure3}}. The design is imported into a finite-element simulation software, Fastcap, which yields the capacitance matrix of the system. The system includes a single-electron pump dot, a counting island, and a detector. The capacitance matrix describes how much gate charge appears on the capacitors due to a sets of gate voltages. The extracted mutual and total capacitances therefore determines the charge sensitivity of the detector \cite{Rossi2010}.
Based on the capacitance matrix, we simulate the charge sensing properties of the superconducting single-electron transistor with a commercial software \mbox{SIMON} \cite{simon}. 
\newpage
The design is changed to optimize the capacitive coupling between the detector and the CI, and the charging energy of the CI and the detector. A design that is similar to the device used in this experiment predicts a charge event at the CI to induce a detector gate charge of \mbox{${\delta q_{\mathrm{e}}} = 2 \times 10^{-2}~e$} in the standard-coupling mode. However, in the enhanced-coupling mode, the single-charge sensitivity increases to \mbox{${\delta q_{\mathrm{e}}} = 2.15 \times 10^{-1}~e$}. The simulated total charging energy of the counting island is $E_{\rm{c}} = 290~\mu$eV.

\begin{figure}[ht]
\centering
\includegraphics[width=\linewidth]{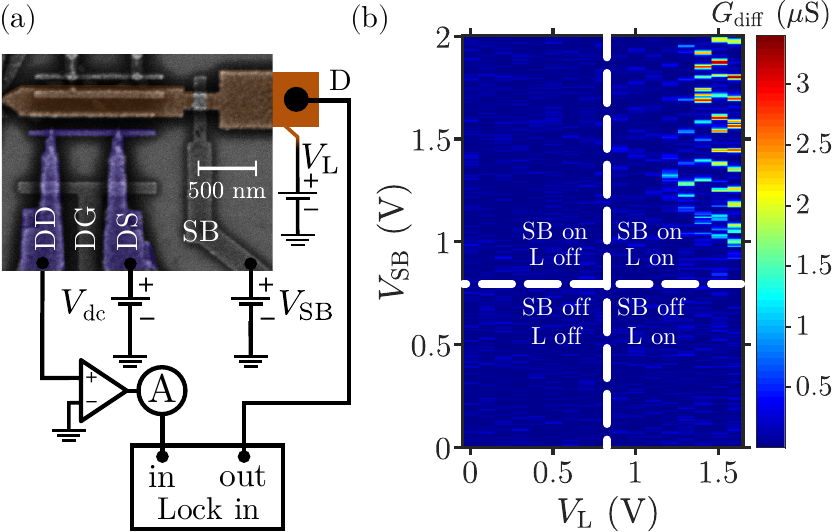}
\caption{(a) Measurement scheme for extracting the capacitance between the detector and counting island (b) Differential conductance corresponding to (a) as a function of the gate voltages $V_{\rm{L}}$ and $V_{\rm{SB}}$. The dashed lines indicate the four gate voltage responses.}
\label{fig:appfigure1}
\end{figure}

\section{Extraction of capacitive coupling}
\label{appendix:coupling}
We use a lockin amplifier to measure the capacitive coupling between the 2DEG channel and the SSET in the standard-coupling mode presented in \mbox{Fig.~\ref{fig:appfigure1}(a)}. Owing to the presence of the finite coupling between the detector and the L and SB gates, also the dc current has to be monitored to extract the detector sensitivity. \mbox{Figure~\ref{fig:appfigure1}(b)} shows the conductance between the 2DEG and the detector, $G = I_{\rm{ac}}^{\rm{rms}}/V_{\mathrm{ac}}^{\mathrm{rms}}$, where  $V_{\mathrm{ac}}^{\mathrm{rms}}$ and $I_{\rm{ac}}^{\rm{rms}}$ are the rms voltage excitation from the lockin amplifier and the measured rms ac current, respectively. The four sections of the plot present the four possible channel configurations. When one of the gates is below the threshold voltage, the conductance is negligible, and hence the ac excitation is blocked. 
In the section where both gates are above the threshold voltage, periodic conductance lines appear, which show periodicity that is caused by changing value of dc voltages $V_{\rm{SB}}$ and $V_{\rm{L}}$.

To estimate the coupling based on \mbox{Eq.~\eqref{eq:coupling}}, first the $n^{\mathrm{rms}}$ gate charge has to be extracted. At a certain $V_{\rm{SB}}$ and $V_{\rm{L}}$ voltages, we record $I_{\rm{dc}}$ and $I^{\rm{rms}}_{\rm{ac}}$ simultaneously, which provide the rms range for the current-to-charge conversion  $[I_{\rm{dc}}-I^{\rm{rms}}_{\rm{ac}}, I_{\rm{dc}}+I^{\rm{rms}}_{\rm{ac}}]$ that results a gate charge signal $[n_{\rm{g,1}},n_{\rm{g,2}}]$  with the help of the calibration experiment shown in \mbox{Fig.~\ref{fig:figure2}(b)}. Because the ac excitation amplitude $V_{\mathrm{ac}}^{\mathrm{rms}}$ is 4 orders of magnitude smaller than the gate charge periodicity induced by the SB, we use the approximation $n^{\mathrm{rms}} = \lvert n_{\rm{g,2}}-n_{\rm{g,1}} \rvert/2$. Thus we may use the known value for the applied $V^{\mathrm{rms}}$ to obtain the coupling capacitance from \mbox{Eq.~\eqref{eq:coupling}}.

\begin{figure}[ht]
\centering
\includegraphics[width=\linewidth]{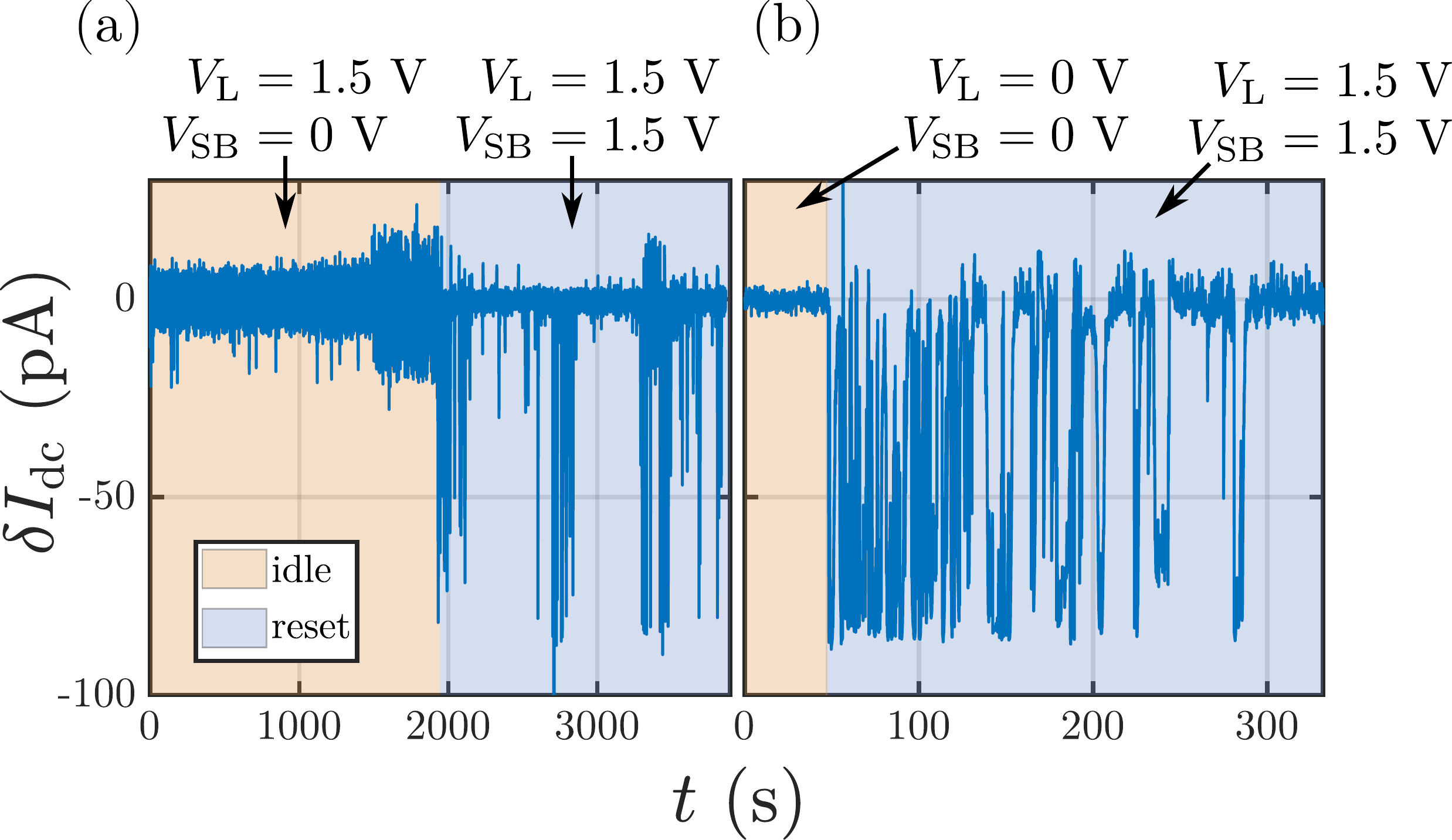}
\caption{Charge jump signal in the enhanced-coupling mode as a function of time (a) The detector noise, when the CI is decoupled from the drain (red) and switched to reset (blue). The effective bath temperature is $\Tbath = 300$ mK. (b) The detector noise, when no CI is present (red), and when we have created a shunting 2DEG under the gate L (blue). The effective bath temperature is $\Tbath = 415 $ mK.}
\label{fig:appfigure2}
\end{figure}

\section{Time traces of random jumps}
\label{appendix:jumps}
The SSET in the enhanced-coupling mode exhibits prominent transient jump noise which we observe in the measurements. Since the jumps reach, in terms of current, the boundaries of the current-to-charge conversion, the mean charge state of the detector island varies. The raw detector signal in the vicinity of the reset is depicted in \mbox{Fig.~\ref{fig:appfigure2}}. 
In Fig.~\ref{fig:appfigure2}(a), we have a floating CI, \mbox{$V_{\rm{L}} = 1.5 $ V}, and monitor the jumps which are used to produce the statistics presented in Fig. \ref{fig:figure3}(b). At \mbox{$t = 1500 $ s}, the noise increases in a step-like manner as observed in \mbox{Sec.~\ref{subsect:change_noise}}, which is followed by a constant noise level. The CI reset, i.e., \mbox{$V_{\rm{L}}$} is set to 1.5 V, starts at \mbox{$t = 1935 $ s} that changes the potential landscape of the system and activates charge relocation not only in the 2DEG but also in the possible local charge traps. However, the white-noise level decreases by 80\% compared with the idle state. This decrease in the noise is attributed to the shunted 2DEG that is shunting the noise subject to the detector island.

In Fig.~\ref{fig:appfigure2}(b), the CI is created simultaneously with the reset. Initially, the lack of CI results a comparable noise with the shielded detector presented in Fig.~\ref{fig:appfigure2}(a). As we apply $1.5 $ V to both L and SB gates, similar jumps appear as in Fig. \ref{fig:appfigure2}(a) on the detector.

\bibliography{My_Collection}
\end{document}